\documentclass[conference]{IEEEtran}
\IEEEoverridecommandlockouts
\makeatletter
\def\ps@headings{%
\def\@oddhead{\mbox{}\scriptsize\rightmark \hfil \thepage}%
\def\@evenhead{\scriptsize\thepage \hfil \leftmark\mbox{}}%
\def\@oddfoot{}%
\def\@evenfoot{}}
\makeatother
\usepackage{subfigure}
\usepackage{colortbl}
\usepackage{bm}

\usepackage{graphicx}  
\usepackage{url}       

\usepackage{amsmath}   
\usepackage{cite}

\usepackage{amsfonts,amssymb}

\usepackage{stfloats}

\usepackage{cases}
\usepackage{algorithm}
\usepackage{multirow}
\usepackage{algorithmic}
\usepackage{stfloats}
\usepackage{epstopdf}

\makeatletter

\newcommand{\Rmnum}[1]{\expandafter\@slowromancap\romannumeral #1@}
\makeatother

\usepackage[absolute,showboxes]{textpos}

\TPMargin{3pt}

\newcommand{\copyrightstatement}{
    \begin{textblock}{0.84}(0.08,0.93) 
         \noindent
         \footnotesize
         \copyright 2022 IEEE. Personal use of this material is permitted. Permission from IEEE must be obtained for all other uses, in any current or future media, including reprinting/republishing this material for advertising or promotional purposes, creating new collective works, for resale or redistribution to servers or lists, or reuse of any copyrighted component of this work in other works. DOI: 10.1109/MCOM.001.2100704.
    \end{textblock}
}

\newcommand{\ls}[1]
    {\dimen0=\fontdimen6\the\font
     \lineskip=#1\dimen0
     \advance\lineskip.5\fontdimen5\the\font
     \advance\lineskip-\dimen0
     \lineskiplimit=.9\lineskip
     \baselineskip=\lineskip
     \advance\baselineskip\dimen0
     \normallineskip\lineskip
     \normallineskiplimit\lineskiplimit
     \normalbaselineskip\baselineskip
     \ignorespaces
    }

\begin{document}
\copyrightstatement

\title{OAM-SWIPT for IoE-Driven 6G}
\vspace{10pt}
\author{\IEEEauthorblockN{Runyu Lyu, Wenchi Cheng, Bazhong Shen, Zhiyuan Ren, and Hailin Zhang}\\[0.2cm]
\vspace{-10pt}

%
\vspace{-30pt}

\thanks{This work was supported in part by the National Key Research and Development Program of China under Grant 2021YFC3002102.}

}

\maketitle

\begin{abstract}
Simultaneous wireless information and power transfer (SWIPT), which achieves both wireless energy transfer (WET) and information transfer, is an attractive technique for future Internet of Everything (IoE) in the sixth-generation (6G) mobile communications. With SWIPT, battery-less IoE devices can be powered while communicating with other devices. Line-of-sight (LOS) RF transmission and near-field inductive coupling based transmission are typical SWIPT scenarios, which are both LOS channels and without enough degree of freedom for high spectrum efficiency as well as high energy efficiency. Due to the orthogonal wavefronts, orbital angular momentum (OAM) can facilitate the SWIPT in LOS channels. In this article, we introduce the OAM-based SWIPT as well as discuss some basic advantages and challenges for it. After introducing the OAM-based SWIPT for IoE, we first propose an OAM-based SWIPT system model with the OAM-modes assisted dynamic power splitting (DPS). Then, four basic advantages regarding the OAM-based SWIPT are reviewed with some numerical analyses for further demonstrating the advantages. Next, four challenges regarding integrating OAM into SWIPT and possible solutions are discussed. OAM technology provides multiple orthogonal streams to increase both spectrum and energy efficiencies for SWIPT, thus creating many opportunities for future WET and SWIPT researches.
\end{abstract}

\vspace{5pt}

\begin{IEEEkeywords}
Wireless energy transfer (WET), simultaneous wireless information and power transfer (SWIPT), orbital angular momentum (OAM), reconfigurable intelligent surface (RIS).
\end{IEEEkeywords}

\section{Introduction}
\IEEEPARstart{W}{ith} the fifth-generation (5G) mobile communication networks eventually deployed, it is urgent to develop the sixth-generation (6G) mobile communication technologies. One of the most expecting research areas of 6G is the Internet of Everything (IoE)\cite{6G}, which denotes the intelligent connection among people, processes, data, and things. Integrated with cloud computing and big data, IoE can connect ``everything'' online, thus achieving the overall planning for real-time data processing, resource allocation, and quality of service (QoS) adjustment. IoE also facilitates many 6G driving applications, including autonomous cars, autonomous drone swarms\cite{Yao_5G_UAV_ResourceAllocation}, smart cities, and smart agriculture. Thus, the IoE-driven 6G is an important research direction for 6G. Because IoE connects billions of passive devices, how to provide energy for these IoE devices is very challenging. Harvesting energy from the surrounding environment is a sustainable way to prolong the lifetime of IoE devices. Apart from natural energy sources such as solar, wind, tide, and heat, the radio-frequency (RF) signal can also be available. In addition, using RF signals carried with information for wireless energy transfer (WET) can realize simultaneous wireless information and power transfer (SWIPT)\cite{SWIPT,SWIPT_Kun}, which can significantly improve the performance of information networks in terms of spectrum efficiency and energy efficiency.

However, there are two problems limiting the spectrum efficiency and energy efficiency of SWIPT. The first problem is the low degree of freedom for SWIPT channels. {For powering the massive IoE devices and facilitating the interference cancellation, the energy source of WET should be stable and fully controllable over power, waveforms, and time/frequency dimensions\cite{WPC}. Therefore, line-of-sight (LOS) RF transmission and near-field inductive coupling based transmission would be the typical SWIPT scenarios}. {Although the direct path in the LOS scenario seemingly facilitates the signal-to-noise ratio (SNR) at the receiver, the strong channel correlation of the LOS channel severely limits the degree of freedom and capacity gain when using traditional multiple antennas for relatively high frequency bands.} Fortunately, orbital angular momentum (OAM)\cite{oam_light}, which is the angular momentum of electromagnetic waves around the propagation axis, is a promising technology for 6G ecosystem. Due to the orthogonal wavefront of OAM carried electromagnetic waves with different topological charges (also OAM-modes), OAM waves can be multiplexed/demultiplexed together and increase the spectrum efficiency without additional power, thus increasing the energy efficiency and facilitating SWIPT in LOS channels based scenarios. The feasibility of using OAM for multiplexing signals transmission has been validated by studies and experiments\cite{oam_low_freq_radio}. However, few studies considered using OAM for WET or SWIPT\cite{SWIPT_OAM2}, 
not to mention discussing the challenges of integrating OAM into SWIPT.

\begin{figure*}[htbp]
\centering
\includegraphics[scale=0.27]{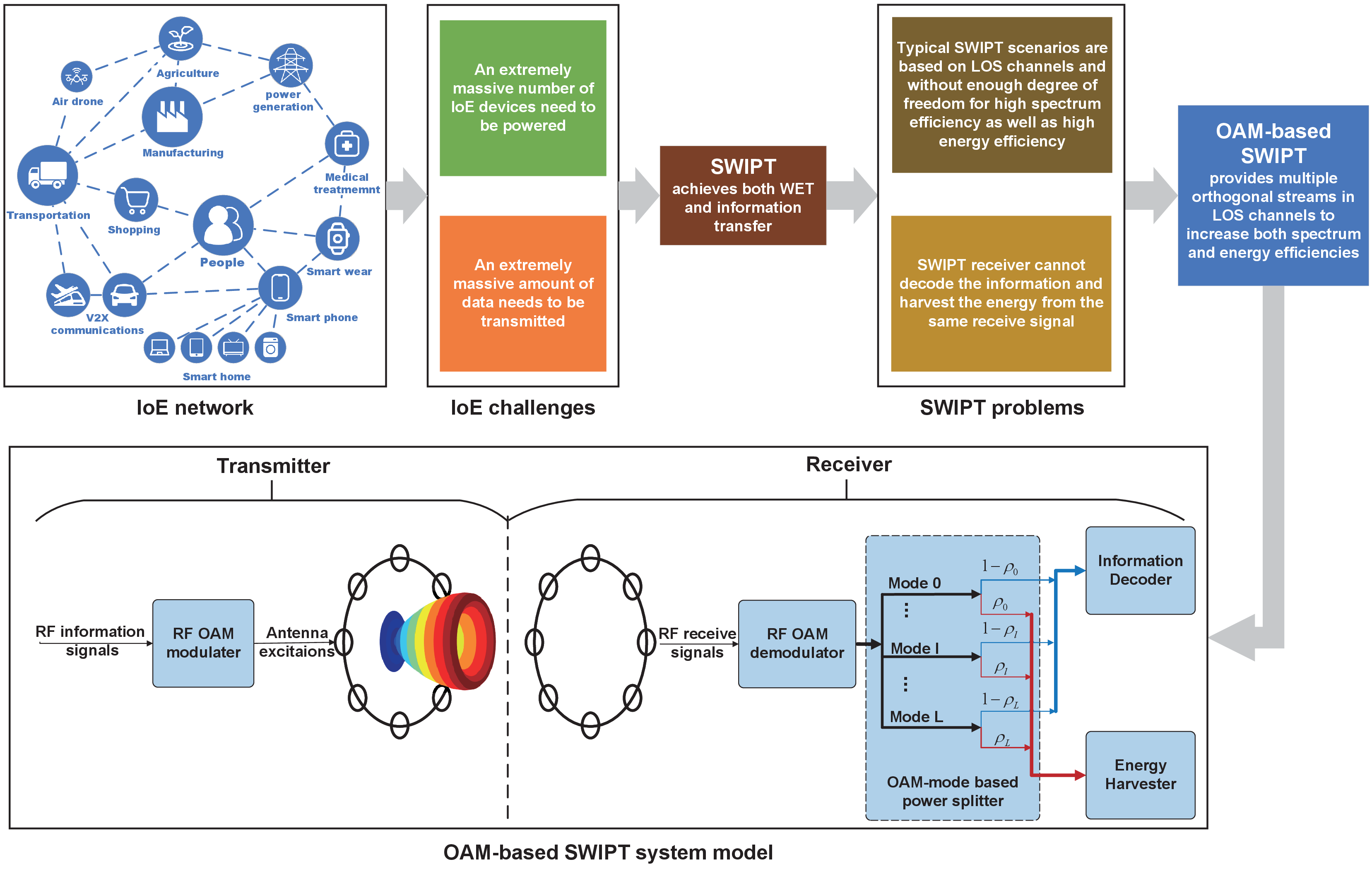}
\caption{OAM-based SWIPT system model for IoE.} \label{fig:IoE2OAM}
\vspace{-15pt}
\end{figure*}

Another problem for SWIPT is that the SWIPT receiver cannot decode the information and harvest the energy from the same receive signal. This is because practical circuits that harvest energy from RF signals cannot decode the information carried by the signals.
Therefore, the receive energy needs to be allocated for information decoding (ID) and energy harvesting (EH), which introduces a tradeoff between the transmission rate and energy efficiency. To allocate the receive energy, the authors of \cite{SWIPT_MIMO} proposed the time switching (TS) and power splitting (PS) schemes. For the TS scheme, the receiver switches between ID and EH mode in a time-division manner. For the PS scheme, the receive signals {are split} into two streams with different power levels for ID and EH. The TS-based and PS-based SWIPT schemes have been studied in various SWIPT systems such as the multiple-input-multiple-output (MIMO) system\cite{SWIPT_MIMO} and the magnetic-induction system\cite{WJY_MI_SWIPT}. The authors of \cite{SWIPT_OPS} generalize the TS and PS schemes to a general scheme called dynamic power splitting (DPS), by which the signals can be dynamically split into two streams for ID and EH. The DPS scheme improves the flexibility for the allocation of receive energy, thus making it more possible to find different optimal allocation schemes for different SWIPT scenarios than TS and PS schemes. Therefore, combining SWIPT with OAM and DPS can achieve high efficient and adaptable energy-information transfer for 6G. However, the main challenges of integrating OAM into SWIPT include the low energy efficiency for OAM-based WET caused by severe divergence of OAM waves, the susceptibility of OAM to transceiver misalignment, the future application of OAM-based SWIPT in non-line-of-sight (NLOS) scenarios, and the mode selection for multi-user OAM-based SWIPT.


In this article, we introduce the OAM-based SWIPT and discuss its basic advantages and challenges. We first introduce the OAM-based SWIPT for IoE. Next, we propose an OAM-based SWIPT system model using OAM-modes assisted DPS scheme, which can dynamically allocate the receive energy of each OAM-mode signal for ID and EH. Multiple orthogonal streams of OAM waves make OAM-based SWIPT more energy efficient and spectrum efficient than traditional single-input-single-output (SISO) and LOS-MIMO based SWIPTs. Then, four basic advantages regarding the OAM-based SWIPT are reviewed, including high spectrum efficiency, supporting multi-user SWIPT, flexible energy allocation, and wide range of versatility. We also give some numerical analyses to further demonstrate the advantages. Four challenges regarding integrating OAM into SWIPT are discussed finally, including low energy efficiency for high-order OAM-based WET, strict requirements for transceiver alignment, OAM-based SWIPT in NLOS channels, and OAM-based multi-user mode selection, along with potential research directions.

The rest of the paper is organized as follows. Section~\ref{sec:IoE2OAM} introduces the OAM-based SWIPT for IoE. In Section~\ref{sec:IoE2OAM}, we also give an OAM-based SWIPT system model using DPS scheme. Section~\ref{sec:advantage} discusses four basic advantages of OAM-based SWIPT. In Section~\ref{sec:challenge}, we discuss four challenges regarding integrating OAM into SWIPT along with possible solutions. Finally, the conclusion is given in Section~\ref{sec:Conclusion}.

\section{OAM-based SWIPT System model for IoE}\label{sec:IoE2OAM}
One of the key features for 6G is its extremely massive connectivity for small devices to enable IoE, which refers to the intelligent connection among people, processes, data, and things. IoE evolves from the Internet of things (IoT), which is a giant information network combining various IoT devices, such as RF identification (RFID) devices, infrared sensors, global positioning systems, and laser scanners. With the development of the Internet, more and more small devices are connected and controlled by the Internet, making the amount of IoT devices and interactive data grow sharply, thus evolving IoT into IoE. Fig.~\ref{fig:IoE2OAM} OAM-based SWIPT system model for IoE shows that massive IoE devices, such as smart wear, smart home appliances, RFID logistics, medical monitors, and agricultural growth monitors, are connected by the IoE network. IoE achieves the overall planning for real-time data processing, resource allocation, and QoS adjustment, thus facilitating many 6G driving applications, such as autonomous cars, autonomous drone swarms, smart cities, and smart agriculture. However, there are two main challenges for IoE. First, an extremely massive number of IoE devices need to be powered. Second, an extremely massive amount of data needs to be transmitted. SWIPT can achieve both WET and information transfer, making it a promising solution to these challenges. For the sake of energy efficiency, LOS channels will be typical SWIPT scenarios, where the low degree of freedom severely decreases the spectrum efficiency. Also, because SWIPT receiver cannot decode the information and harvest the energy from the same receive signal, the energy needs to be allocated for ID and EH. OAM can facilitate high efficient SWIPT in LOS channels because it provides multiple orthogonal streams in LOS channels to increase both spectrum and energy efficiencies.


To demonstrate our proposed OAM-based SWIPT more clearly, we also give a DPS-assisted OAM-based SWIPT system model in Fig.~\ref{fig:IoE2OAM}. For the transmitter, the multi-stream RF information signals are first modulated into the OAM excitations to antennas. The excitations are with equal amplitudes and linearly increasing phases of $2\pi l/N_t$ for different OAM-modes, where $N_t$ denotes the number of antennas in the transmitter and $l$ represents the index of OAM-mode ($0\le l\le N_t-1$). Then, the excitations of multiple OAM-modes are emitted via the transmitter and form the OAM waves. In this article, we chose uniform circular arrays (UCAs) as the transmit and receive antennas. This is because UCA can generate waves with multiple OAM-modes digitally and simultaneously.
For the OAM-based-SWIPT receiver, the RF receive signals are first demodulated into the signals corresponding to different OAM-modes. {Then, each OAM-mode signal splits into the ID stream and the EH stream. One OAM-mode signal is shared by ID and EH streams. The power ratio for ID stream to EH stream of the $l$th OAM-mode is set as $\rho_l$ to $1-\rho_l$, where $0\le \rho_l \le 1$.} Since there are totally {$N_t$} OAM-modes, the power splitting ratio can be denoted by a vector as $\boldsymbol{\rho}=[\rho_0,\rho_1,\cdots,\rho_{N_t-1}]$. By choosing specific values of $\boldsymbol{\rho}$, maximum data rate can be achieved under the minimum EH constraint. $\boldsymbol{\rho}$, maximum data rate can be achieved under the minimum EH constraint. {Combined with the reconfigurable intelligent surface (RIS), the sum achievable rate subject to the EH requirement can be further increased by optimizing the reflection coefficients and the power split ratio.} This optimization problem can be classified into the weighted sum rate (WSR) maximization problem. However, since the EH constraint is non-convex, it is challenging to solve this problem. In this article, we use Monte Carlo method to simply derive the optimal values of $\boldsymbol{\rho}$ for numerical analyses. {Specifically, we independently generate a large number of uniform random vectors $\boldsymbol{\rho}$'s (100000 random $\boldsymbol{\rho}$'s), so that the maximum rate can be given as the largest rate calculated by all $\boldsymbol{\rho}$'s under the minimum EH constraint.} Finally, the ID and EH streams are sent to the information decoder and energy harvester, respectively.

\section{Advantages and Rate-Energy performances Analysis for OAM-based SWIPT}\label{sec:advantage}
OAM can increase the spectrum efficiency for ID without additional power and bandwidth, thus increasing the energy efficiency for the SWIPT system. Also, different orthogonal OAM-modes can support more SWIPT users as well as improve the flexibility for energy allocation. In addition, the OAM-based SWIPT, which can be used for LOS RF transmission, near-field inductive coupling based transmission, and optical transmission, has a wide range of versatility. These basic advantages of the OAM-based SWIPT are discussed in detail as follows.

\begin{figure*}[htbp]
\centering
\includegraphics[scale=0.47]{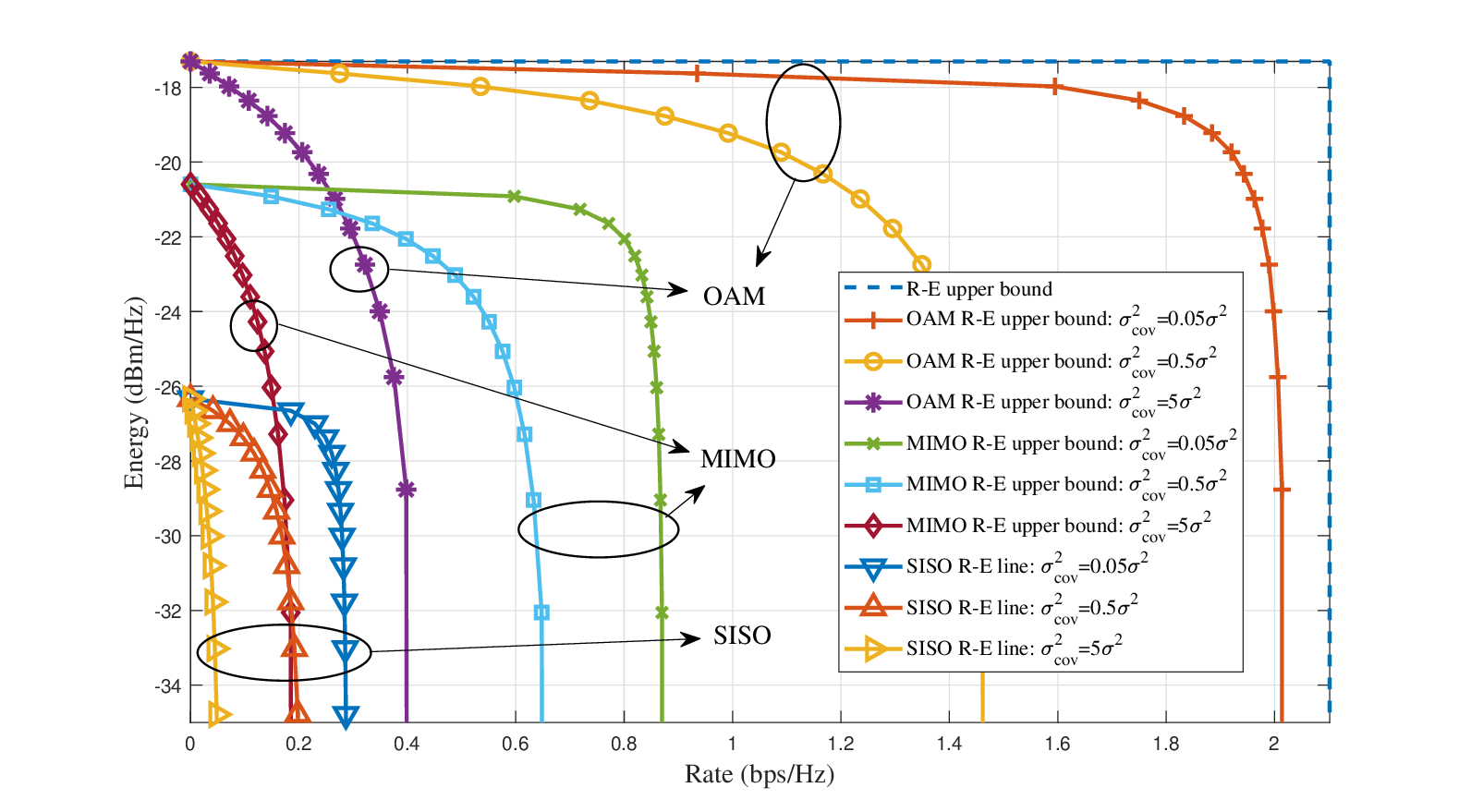}
\caption{{Rate-energy tradeoff for OAM versus MIMO versus SISO with the ratio between the noises of channel ($\sigma^2$) and RF-to-baseband conversion ($\sigma_{cov}^2$).}} \label{fig:RE_upper_sigmacov}
\vspace{-10pt}
\end{figure*}
\begin{figure*}[htbp]
\centering
\includegraphics[scale=0.47]{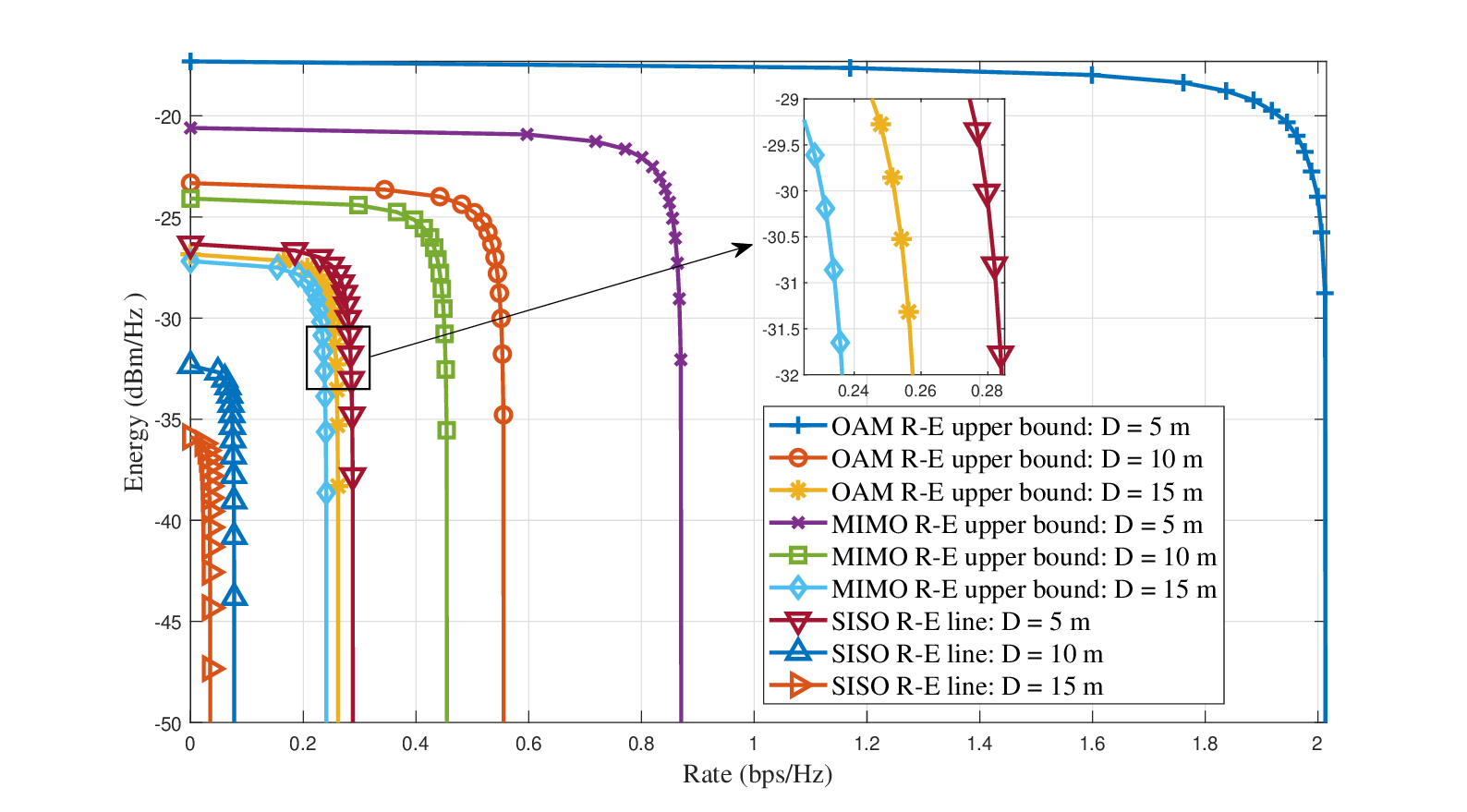}
\caption{{Rate-energy tradeoff for OAM versus MIMO versus SISO with different distances between the transceivers.}} \label{fig:RE_upper_D}
\vspace{-15pt}
\end{figure*}
\textbf{Advantage 1: High spectrum efficiency: }Different OAM-modes are orthogonal with each other, thus creating parallel subchannels for the information transmission. With orthogonality, parallel subchannels can be used for transmitting multi-stream information simultaneously without consuming traditional orthogonal resources such as frequency, code, or space. Also, jointly used with traditional orthogonal resources, OAM can increase the spectrum efficiency for information transfers more significantly. In addition, compared with MIMO technology, OAM can be used in correlated channels such as the LOS channel, which is the typical channel for SWIPT and without enough degree of freedom. The high orthogonality of different OAM-modes in LOS channel increases the potential value of OAM in SWIPT systems as well as other 6G network systems.

\textbf{Advantage 2: Supporting more SWIPT users: }Orthogonal OAM-modes can be utilized for multi-user SWIPT. Considering the extraordinary amount of IoE devices in future 6G networks, it is highly needed to support multi-user SWIPT. As a new orthogonal resource, OAM can achieve mode division multiple access (MDMA) for both WET and information transfer. {For multiple users that are coaxial with each other, different OAM-modes are orthogonal with each other and can be used to support multiple users. Also, OAM can be combined with non-orthogonal multiple access (NOMA) to further increase the sum-capacity of the near and the far users with acceptable receiving complexity. If UCAs are not perfectly aligned, the proposed beam steering scheme in \cite{OAM_beam_steering_aligned} can be employed such that almost no performance degradation exhibits with respect to the perfectly aligned UCAs.} In addition, jointly used with beamforming technology or other orthogonal resources, the number of devices that OAM-based SWIPT can support will be further increased.

\textbf{Advantage 3: Flexible energy allocation: }Orthogonal OAM-modes also provide more parallel streams for the energy allocation of ID and EH. Combining with DPS scheme, OAM-based SWIPT can achieve more flexible energy allocation than SISO and LOS-MIMO based SWIPTs. Thus, different optimal allocations can be used for different scenarios. Furthermore, combined with cognitive communication and machine learning, OAM-based SWIPT can change the energy allocation levels for ID and EH according to the environment in real time, making it available for various time-variant channels.

\textbf{Advantage 4: Wide range of versatility: }The RF WET scenarios can be classified into three cases, including near-field WET employing inductive coupling, far-field directive power beaming, and far-field ambient RF power scavenging. OAM technology can be used in all of the scenarios to improve the rate-energy (R-E) performances. For near-field inductive coupling based SWIPT, OAM technology can significantly increase the spectrum efficiency\cite{OAM_NFC} as well as provide multi streams for DPS of EH and ID. For far-field directive power beaming, which are often in LOS channels, OAM technology can increase the degree of freedom and improve the R-E performances as discussed in \textbf{Advantage 1} and \textbf{Advantage 2}. In the last case, OAM technology can also increase the directionality of the beams. In addition, OAM can also be used in optical scenarios.

\begin{figure}[htbp]
\centering
\includegraphics[scale=0.105]{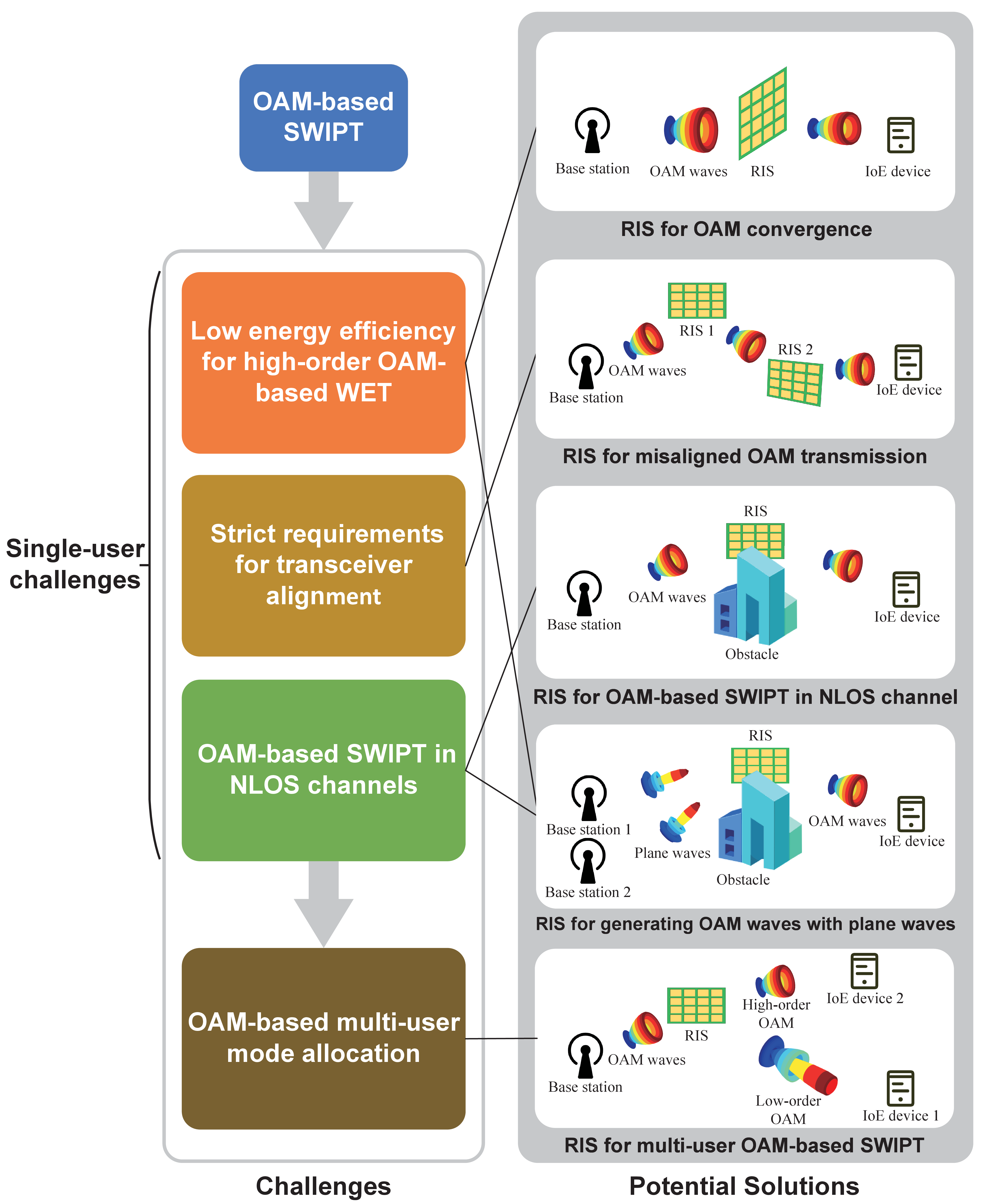}
\caption{Challenges and solutions for OAM-based SWIPT.} \label{fig:challenges_solutions}
\vspace{-15pt}
\end{figure}
\begin{figure*}[hbp]
\centering
\includegraphics[scale=0.32]{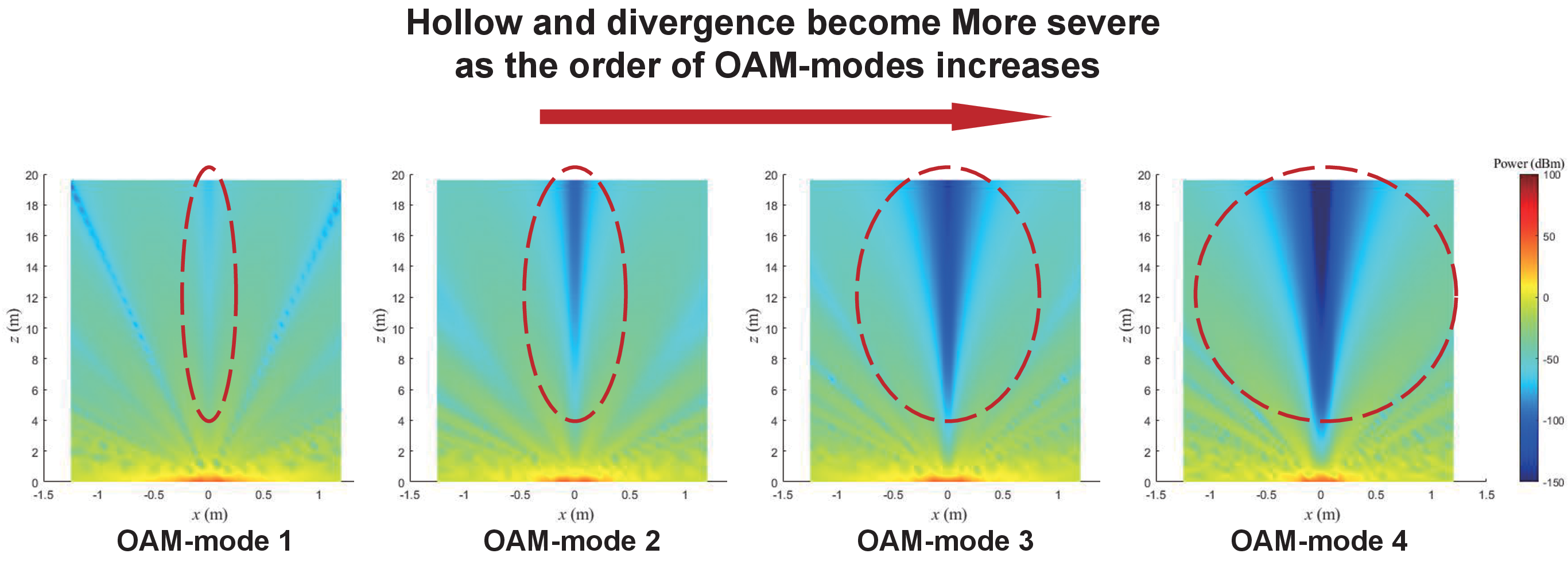}
\caption{Hollow and divergent OAM waves.} \label{fig:OAM_hollow}
\vspace{-15pt}
\end{figure*}
\textbf{Numerical example: }To further demonstrate the advantages of the OAM-based SWIPT, we give numerical analyses for the R-E performances of OAM-based SWIPT by a so-called R-E region in Figs.~\ref{fig:RE_upper_sigmacov} and \ref{fig:RE_upper_D}, where the $x$ axis represents {the maximum bits that can be transmitted through per second per Hz} and the $y$ axis represents the upper bound of the harvested {power per Hz}. {In the rectenna, the received RF signal is converted into a DC signal by a rectifier, which consists of a Schottky diode and a passive low-pass filter. We set the RF-DC conversion efficiency as $1$ to give the maximum harvesting power since the purpose of this paper is to discuss the advantages and challenges of using OAM for SWIPT.} {The bandwidth used in the numerical example is set as $1$ Hz.} Fig.~\ref{fig:RE_upper_sigmacov} shows the R-E upper bounds of OAM, MIMO, and SISO based SWIPTs in a LOS channel. {The OAM-based and MIMO-based SWIPTs are based on the same pair of UCA transmitter and receiver.} The transmit and receive UCAs are aligned with each other, where the SISO transceivers are located at the center of them, respectively. Both UCAs are with $8$ elements and their radii are set as $0.1$ m. The distance between the transmit and receive UCAs is set as {$5$} m. In Fig.~\ref{fig:RE_upper_sigmacov}, the frequency of the RF signal is set as $28$ GHz with the {total transmit power for all OAM-mode signals set as $40$ dBm/Hz. The power of the additive Gaussian white noise at the receiver, denoted by $\sigma^2$, is set as $-20$ dBm/Hz, and the powers for the noises of RF-to-baseband conversion, denoted by $\sigma_{cov}^2$, is set as $0.05\sigma^2$, $0.5\sigma^2$, and $5\sigma^2$, respectively.} {DPS is also used for the MIMO-based SWIPT to allocate the received energy of each MIMO sub-channel.} Fig.~\ref{fig:RE_upper_sigmacov} shows that the R-E upper bounds for OAM, MIMO, and SISO based SWIPTs decrease as $\sigma_{cov}^2$ increases. {Also, OAM and MIMO-based SWIPTs have much larger R-E upper bounds than the SISO based SWIPT for short-transmission distance.} Moreover, the OAM-based SWIPT enjoys a larger R-E upper bound than the MIMO-based SWIPT with the same $\sigma_{cov}^2$. This is because OAM waves can maintain a higher degree of freedom in the LOS channel compared with MIMO signals, thus being more spectrum efficient. However, the gaps between them decrease as $\sigma_{cov}^2$ increases. Therefore, the OAM-based SWIPT has a better R-E performance than the MIMO and SISO based SWIPTs in LOS scenarios.

Figure~\ref{fig:RE_upper_D} shows the R-E upper bounds of OAM, MIMO, and SISO based SWIPTs, where the distance between the transmit and receive antennas is given as {$5$, $10$, and $15$ m with $\sigma_{cov}^2$ set as $0.05\sigma^2$}. {The values of other variables are the same as Fig.~\ref{fig:RE_upper_sigmacov}}. In Fig.~\ref{fig:RE_upper_D}, the R-E upper bounds of OAM-based, MIMO-based, and SISO-based SWIPT systems increase as the distance decreases. Among them, the R-E upper bound of the OAM-based SWIPT system increases much faster than those of the MIMO-based and SISO-based SWIPT systems. This is because{, as the transmission distance decreases, the divergence of each OAM-mode beam decreases, thus increasing the received energy, which increases the SINR.} Therefore, the OAM-based SWIPT is more suitable for the short-distance LOS transmission than the MIMO-based SWIPT.


\section{Challenges for OAM-based SWIPT}\label{sec:challenge}
Despite all of the above advantages, there are some challenges regarding integrating OAM into SWIPT, including low energy efficiency for high-order OAM-based WET, strict requirements for transceiver alignment, OAM-based SWIPT in NLOS channels, as well as OAM-based multi-user mode selection. In the following, we discuss these challenges and give some potential solutions.

\textbf{Challenge 1: Low energy efficiency for high-order OAM-based WET: }The OAM waves are hollow and divergent, which becomes more severe as the order of OAM-modes increases, resulting in severe attenuation of high-order OAM waves and degrading the energy efficiency of WET. As shown in Fig.~\ref{fig:OAM_hollow}, the divergence becomes more and more severe as the order of OAM-mode increases from $1$ to $4$, the received power also decreases. Therefore, only a limited number of OAM-modes can be directly used for OAM-based SWIPT, which decreases the spectrum efficiency and {also reduces
the maximum energy for harvesting}. For instance, the metasurface and the RIS can generate convergent OAM waves\cite{oam_metasurface2} {by changing the phase, amplitude, and frequency of the incident electromagnetic waves. Compared with the beamforming antenna array,  metasurface and RIS do not need radio frequency chains, thus reducing the complexity}. In the future, other kinds of nonplanar waves, which have orthogonal wavefronts and convergent beams, can {also} be used for energy efficiency enhancement of OAM-based SWIPT.

\begin{figure}[htbp]
\centering
\includegraphics[scale=0.6]{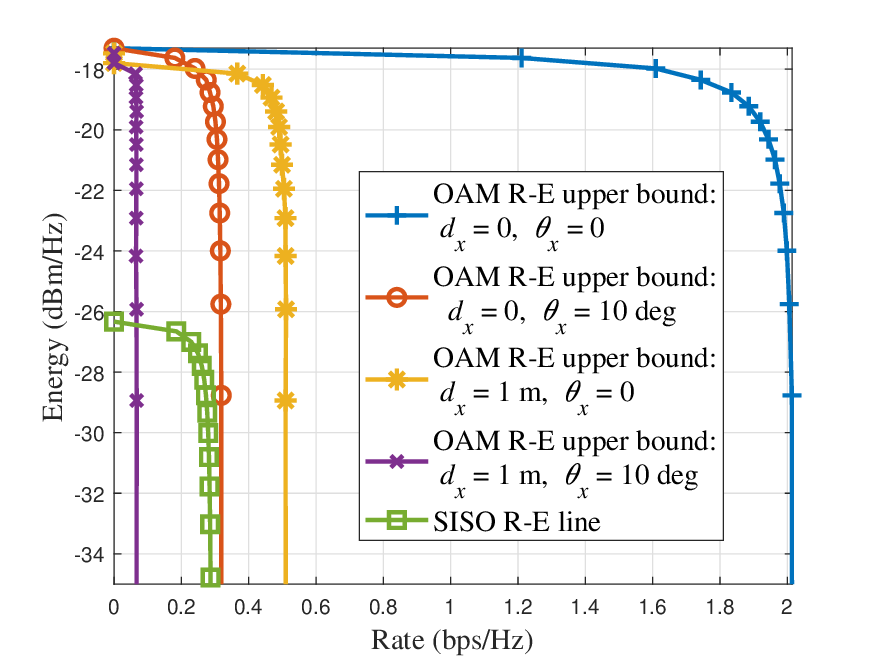}
\caption{{Rate-energy tradeoff for OAM versus SISO with misaligned transceivers.}} \label{fig:RE_upper_dx_thetax}
\vspace{-15pt}
\end{figure}
\textbf{Challenge 2: Strict requirements for transceiver alignment: }The misalignment between the OAM transmitter and receiver can cause distorted OAM waves, which severely reduces the orthogonality among different OAM-modes, leading to a sharp degradation in OAM-based SWIPT R-E performance. To demonstrate the degradation brought by transceiver misalignment, we give numerical analysis for the R-E performance of the OAM-based SWIPT with misaligned transceivers in Fig.~\ref{fig:RE_upper_dx_thetax}, where $\theta_x$ denotes the angle between the normal lines of the transmit and receive UCAs, $d_x$ denotes the distance between the normal lines. $\sigma_{cov}^2$ is set as {$0.05\sigma^2$} and other variables {are the same as} Fig.~\ref{fig:RE_upper_sigmacov}. As shown in Fig.~\ref{fig:RE_upper_dx_thetax}, the R-E upper bound of OAM-based SWIPT severely decreases as $\theta_x$ and $d_x$ increase. Furthermore, for OAM-based SWIPTs with $d_x = 1$ m {and $\theta_x = 10$ deg}, the rate upper bounds are smaller than that of the SISO based SWIPT, which means the transceiver misalignment can offset the spectrum enhancement OAM brings. Some schemes have been developed to achieve non-aligned reception at the cost of increasing the complexity and energy consumption. For example, the channel state information can be obtained with channel estimation {to separate interfered OAM-mode signals\cite{OAM_beam_steering_aligned}}. However, the heavy overhead for channel estimation makes the complexity sharply increased. One of the most promising solutions for energy efficient non-aligned OAM reception is the RIS. As shown in Fig.~\ref{fig:challenges_solutions}, with multiple RIS's, misaligned OAM waves can be reflected and distorted OAM waves can be reconstructed {by programming the reflection amplitude and phase of each RIS element\cite{RIS_multiuser}}. Moreover, because RIS can achieve realtime and flexible control of electromagnetic parameters, such as amplitude, phase, and frequency, the direction and phase structure of incident OAM waves can be quickly changed in real time to react to the time-variant channel. {However, the complexity of the algorithm for optimizing the phase shifts of RIS elements is relatively high, especially for large-size RIS.}


\textbf{Challenge 3: OAM-based SWIPT in NLOS channels: }Although SWIPT scenarios are often with LOS channels, there will be many NLOS scenarios in future IoE networks. Also, because OAM multiplexing is based on the orthogonal wavefronts of different OAM-modes, a LOS path is necessary for maintaining the structures of OAM waves. Thus, it is needed to generate LOS paths for the OAM-based SWIPTs in NLOS scenarios. RIS can reflect the OAM waves blocked by obstacles and construct a LOS path for the OAM-based SWIPT, as shown in Fig.~\ref{fig:challenges_solutions}. Since the RIS is often passive, it is much more power-efficient {and has lower hardware complexity} than conventional active relays. Also, because RIS can achieve complex electromagnetic beamforming, it can reconstruct the distorted OAM waves. In addition, RIS can also reflect plane waves and generate convergent OAM waves, as shown in Fig.~\ref{fig:challenges_solutions}.

\textbf{Challenge 4: Mode selection for multi-user OAM-based SWIPT: }Although OAM can achieve multi-user SWIPT, different IoE devices have various locations, distances, and energy requirements, making OAM-based multi-user SWIPT very challenging. Thus, multi-user mode selection schemes need to be developed for OAM-based multi-user SWIPT. For example, in veiw of the severe divergence of high-order OAM-modes, low-order OAM-modes should be used for high power consumption and long-distance devices. Also, multiple OAM-modes can align to different IoE devices with the help of {RIS\cite{RIS_multiuser}}. {As shown in Fig.~\ref{fig:challenges_solutions}, a RIS, which is deployed between the base station and multiple IoE devices, reflects the incident OAM beam from the base station and directly projects it to the IoE devices at different directions by changing the phase and amplitude of the OAM beam.}

\section{Conclusions}\label{sec:Conclusion}
In this article, we introduced and discussed the advantages and challenges of the OAM-based SWIPT. We first introduced the OAM-based SWIPT for IoE. Next, we proposed a DPS-assisted OAM-based SWIPT system model. Then, four basic advantages regarding the OAM-based SWIPT were reviewed, including high spectrum efficiency, supporting multi-user SWIPT, flexible energy allocation, and wide range of versatility. Despite all of the above advantages, we discussed four challenges regarding integrating OAM into SWIPT, including low energy efficiency for high-order OAM-based WET, strict requirements for transceiver alignment, OAM-based SWIPT in NLOS channels, and OAM-based multi-user mode selection, along with several possible solutions, such as using RIS for OAM-based SWIPT in NLOS scenarios. In conclusion, OAM technology provides multiple orthogonal streams to increase both spectrum and energy efficiencies for SWIPT, creating many opportunities for future WET and SWIPT researches.

\bibliographystyle{IEEEtran}
\bibliography{References}

\begin{thebibliography}{10}
\providecommand{\url}[1]{#1}
\csname url@samestyle\endcsname
\providecommand{\newblock}{\relax}
\providecommand{\bibinfo}[2]{#2}
\providecommand{\BIBentrySTDinterwordspacing}{\spaceskip=0pt\relax}
\providecommand{\BIBentryALTinterwordstretchfactor}{4}
\providecommand{\BIBentryALTinterwordspacing}{\spaceskip=\fontdimen2\font plus
\BIBentryALTinterwordstretchfactor\fontdimen3\font minus
  \fontdimen4\font\relax}
\providecommand{\BIBforeignlanguage}[2]{{%
\expandafter\ifx\csname l@#1\endcsname\relax
\typeout{** WARNING: IEEEtran.bst: No hyphenation pattern has been}%
\typeout{** loaded for the language `#1'. Using the pattern for}%
\typeout{** the default language instead.}%
\else
\language=\csname l@#1\endcsname
\fi
#2}}
\providecommand{\BIBdecl}{\relax}
\BIBdecl

\bibitem{6G}
W.~Saad, M.~Bennis, and M.~Chen, ``A vision of {6G} wireless systems:
  Applications, trends, technologies, and open research problems,'' \emph{IEEE
  Network}, vol.~34, no.~3, pp. 134--142, 2020.

\bibitem{Yao_5G_UAV_ResourceAllocation}
Z.~Yao, W.~Cheng, W.~Zhang, and H.~Zhang, ``Resource allocation for
  {5G-UAV}-based emergency wireless communications,'' \emph{IEEE Journal on
  Selected Areas in Communications}, vol.~39, no.~11, pp. 3395--3410, 2021.

\bibitem{SWIPT}
L.~R. Varshney, ``Transporting information and energy simultaneously,'' in
  \emph{2008 IEEE International Symposium on Information Theory}, 2008, pp.
  1612--1616.

\bibitem{SWIPT_Kun}
J.~Hu, K.~Yang, G.~Wen, and L.~Hanzo, ``Integrated data and energy
  communication network: A comprehensive survey,'' \emph{IEEE Communications
  Surveys Tutorials}, vol.~20, no.~4, pp. 3169--3219, 2018.

\bibitem{WPC}
S.~Bi, C.~K. Ho, and R.~Zhang, ``Wireless powered communication: opportunities
  and challenges,'' \emph{IEEE Communications Magazine}, vol.~53, no.~4, pp.
  117--125, 2015.

\bibitem{oam_light}
L.~Allen, M.~Beijersbergen, R.~Spreeuw, and J.~P.~Woerdman, ``Orbital angular
  momentum of light and transformation of {Laguerre Gaussian} laser modes,''
  \emph{Physical review. A}, vol.~45, pp. 8185--8189, July 1992.

\bibitem{oam_low_freq_radio}
B.~Thide, H.~Then, J.~Sj\"oholm, K.~Palmer, J.~Bergman, T.~D~Carozzi,
  Y.~Istomin, N.~Ibragimov, and R.~Khamitova, ``Utilization of photon orbital
  angular momentum in the low-frequency radio domain,'' \emph{Physical review
  letters}, vol.~99, pp. 087\,701--087\,701, Aug. 2007.

\bibitem{SWIPT_OAM2}
M.~Mase, N.~Shinohara, T.~Mitani, and S.~Ishino, ``Evaluation of efficiency and
  isolation in wireless power transmission using orbital angular momentum
  modes,'' in \emph{2021 IEEE Wireless Power Transfer Conference (WPTC)}, 2021,
  pp. 1--4.

\bibitem{SWIPT_MIMO}
R.~Zhang and C.~K. Ho, ``{MIMO} broadcasting for simultaneous wireless
  information and power transfer,'' \emph{IEEE Transactions on Wireless
  Communications}, vol.~12, no.~5, pp. 1989--2001, 2013.

\bibitem{WJY_MI_SWIPT}
J.~Wang, W.~Cheng, W.~Zhang, W.~Zhang, and H.~Zhang, ``Multi-frequency access
  for magnetic induction based {SWIPT},'' \emph{IEEE Journal on Selected Areas
  in Communications}, pp. 1--1, 2022.

\bibitem{SWIPT_OPS}
X.~Zhou, R.~Zhang, and C.~K. Ho, ``Wireless information and power transfer:
  Architecture design and rate-energy tradeoff,'' \emph{IEEE Transactions on
  Communications}, vol.~61, no.~11, pp. 4754--4767, 2013.

\bibitem{OAM_beam_steering_aligned}
R.~Chen, H.~Xu, M.~Moretti, and J.~Li, ``Beam steering for the misalignment in
  {UCA}-based {OAM} communication systems,'' \emph{IEEE Wireless Communications
  Letters}, vol.~7, no.~4, pp. 582--585, 2018.

\bibitem{OAM_NFC}
R.~Lyu, W.~Cheng, and W.~Zhang, ``Modeling and performance analysis of oam-nfc
  systems,'' \emph{IEEE Transactions on Communications}, pp. 1--1, 2021.

\bibitem{oam_metasurface2}
F.~{Qin}, L.~{Wan}, L.~{Li}, H.~{Zhang}, G.~{Wei}, and S.~{Gao}, ``A
  transmission metasurface for generating {OAM} beams,'' \emph{IEEE Antennas
  and Wireless Propagation Letters}, vol.~17, no.~10, pp. 1793--1796, 2018.

\bibitem{RIS_multiuser}
B.~Di, H.~Zhang, L.~Song, Y.~Li, Z.~Han, and H.~V. Poor, ``Hybrid beamforming
  for reconfigurable intelligent surface based multi-user communications:
  Achievable rates with limited discrete phase shifts,'' \emph{IEEE Journal on
  Selected Areas in Communications}, vol.~38, no.~8, pp. 1809--1822, 2020.

\end{thebibliography}

\end{document}